\documentstyle[emulateapj,psfig]{article}
\begin{document}

\newcommand{\etal}{{\it et al.}}
\newcommand{\chandra}{{\it Chandra}}
\newcommand{\hst}{{\it HST}}

\slugcomment{Accepted for publication in the Astrophysical Journal}

\title{A deep search with {\it HST} for late time supernova signatures in the hosts of XRF 011030 and XRF~020427}

\author{
Andrew Levan\altaffilmark{1,2,3},
Sandeep Patel\altaffilmark{1,4},
Chryssa Kouveliotou\altaffilmark{1,5}
Andrew Fruchter\altaffilmark{1,3},
James Rhoads\altaffilmark{3},
Evert Rol\altaffilmark{1,2},
Enrico Ramirez-Ruiz\altaffilmark{1,6,7},
Javier Gorosabel\altaffilmark{3,8},
Jens Hjorth\altaffilmark{9},
Ralph Wijers\altaffilmark{10},
W. Michael Wood-Vasey\altaffilmark{11},
David Bersier\altaffilmark{3},
Alberto Castro-Tirado\altaffilmark{8},
Johan Fynbo\altaffilmark{9},
Brian Jensen\altaffilmark{9},
Elena Pian\altaffilmark{12},
Nial Tanvir\altaffilmark{1,13},
Stephen Thorsett\altaffilmark{14},
Stan Woosley\altaffilmark{1,14}}

\altaffiltext{1}{Institute for Nuclear Theory, University of Washington, Seattle, Washington 98195-1550, USA}
\altaffiltext{2}{Department of Physics and Astronomy, University of Leicester, University Road,
Leicester, LE1 7RH, UK}
\altaffiltext{3}{Space Telescope Science Institute, 3700 San Martin Drive, Baltimore, MD21218, USA}
\altaffiltext{4}{Universities Space Research Association, National Space Science Technology Center, SD-50, 320 Sparkman Drive, Huntsville, AL
35805, USA}
\altaffiltext{5}{NASA/Marshall Space Flight Center, National Space Science Technology Center, SD-50, 320 Sparkman Drive, Huntsville, AL
35805, USA}
\altaffiltext{6}{School of Natural Sciences, Institute for Advanced Study, Einstein Drive, Princeton,
NJ 08540, USA}
\altaffiltext{7}{Chandra Fellow}
\altaffiltext{8}{Instituto de Astrof\'{\i}sica de Andaluc\'{\i}a (IAA-CSIC),
    P.O. Box 03004, E-18080 Granada, Spain.}
\altaffiltext{9}{Astronomical Observatory, University of Copenhagen, Juliane Maries Vej 30, DK-2100, Copenhagen, Denmark}
\altaffiltext{10}{Astronomical Institute ``Anton Pannekoek''and Center for High Energy Astrophysics, University of Amsterdam, Kruislaan 403,
1098 SJ Amsterdam, NL}
\altaffiltext{11}{Lawrence Berkeley National Laboratory, One Cyclotron Road, Mailstop 50R232, Berkeley, CA 94720, USA}
\altaffiltext{12}{INAF, Osservatorio Astronomico di Trieste, Via G.B. Tiepolo 11, I-34131 Trieste, I}
\altaffiltext{13}{Centre for Astrophysics Research, University of Hertfordshire, College Lane, Hatfield AL10 9AB, UK}
\altaffiltext{14}{Department of Astronomy \& Astrophysics, University of California at Santa Cruz, Santa Cruz, CA 95064, USA}

\begin{abstract}
\noindent X-ray Flashes (XRFs) are, like Gamma-Ray Bursts (GRBs), thought to
signal the collapse of massive stars in distant galaxies. 
Many models posit that the
isotropic equivalent energies of XRFs are lower than those for GRBs,
such that they are visible from a reduced range of distances when
compared with GRBs. Here we present the results
of two epoch {\it Hubble Space Telescope} imaging of two XRFs. 
These images taken approximately 45 
and 200 days post burst reveal no evidence for an associated supernova in
either case. Supernovae such as SN~1998bw would have been
visible out to $z \sim 1.5$ in each case, while fainter supernovae
such as SN~2002ap would have been visible to $z \sim 1$. 
If the XRFs lie at such large distances, their energies 
would not fit the observed correlation
between the GRB peak energy and isotropic energy release ($E_{p} \propto E_{iso}^{1/2}$), in which
soft bursts are less energetic. We conclude that, should these XRFs
reside at low redshifts ($z<0.6$), either their line of 
sight is heavily extinguished, or they are associated with extremely faint supernovae, 
or, unlike GRBs, these XRFs do not have temporally coincident supernovae.
\end{abstract}


\section{Introduction}
X-ray Flashes (XRFs) occupy the extreme left (lower energy) wing of
the peak energy distribution of Gamma-Ray Bursts (GRBs). While GRBs
radiate the majority of their energy in $\gamma$-rays ($E_{p}\sim200$
keV, Preece \etal\ 2000), XRFs are characterized by peak energies
below 50 keV and an X-ray fluence in excess of that observed in
$\gamma$-rays.  XRFs were discovered with the {\it BeppoSAX} Wide
Field Cameras (Heise \etal\ 2001a); they constitute approximately 1/3
of the total burst population (Lamb \etal\ 2004).  The physical
mechanisms driving XRFs and their observed differences to GRBs are
some of the key questions in the field today.

Of the many competing theories relating to the origin of XRFs, the
most popular is that XRFs are produced when a classical GRB is
observed ``off-axis'', so that the highly collimated ejecta (i.e., the
highest energies and harder photons) are not seen (Yamazaki, Ioka \&
Nakamura, 2002; Rhoads 2003; Dado et al. 2004). Alternatively, models
invoking either an increase of the baryon load within the fireball
itself (Dermer, Chiang \& B\"ottcher 1999; Ramirez-Ruiz \& Lloyd-Ronning
2002; Meszaros \etal\ 2002; Huang, Dai \& Lu 2002) or low efficiency
(high baryon load) shocks (Zhang \& Meszaros 2002; Barraud et
al. 2003) can also produce XRFs.

One interesting consequence of all of the models above, is that they predict
a distance distribution for XRFs which is foreshortened with respect
to that of GRBs, since XRFs emit a fraction of the energy of the GRB
itself.  This hypothesis seems to be supported by the analysis of GRBs
with known redshifts, whereby the event peak energy
$(E_{p}$) scales as $E_{iso}^{1/2}$, i.e. softer GRBs have lower
isotropic energies (Amati \etal\ 2002; Lloyd-Ronning \& Ramirez-Ruiz
2002). Qualitative arguments on XRF distances were presented by Kouveliotou \etal\ (2004), 
where a comparison of the
energetics of the X-ray afterglows of GRBs and XRFs suggested that were
the latter placed a high redshift (e.g. $z=1.5$), their luminosities would rival
those of the brightest GRBs, suggesting a closer proximity for XRFs.

Qualitative distance arguments have also been presented by
Kouveliotou \etal\ (2004), where a comparison between the energetics
of GRBs and XRFs indicated that if XRFs were placed further than $z=1.5$
their luminosities would exceed those of the brightest GRBs,
suggesting a closer proximity for XRFs.

However, determining the true distances of XRFs has proven to be
exceptionally difficult observationally.  Only two XRFs have so far
been associated with optical afterglows, XRF~020903 (Soderberg \etal\
2004) and XRF~030723 (Fynbo \etal\ 2004). Of these only the former has
a measured spectroscopic redshift ($z=0.25$; Soderberg \etal\ 2004) while
the latter has only  a redshift limit ($z<2.3$) with the suggestion of low
redshift based on the presence of a supernova like component it its late time light curve
(Fynbo \etal\ 2004). A
second possible case, GRB/XRF 031203 is not yet settled, as there is
an ongoing debate on the true nature of this nearby event ($z=0.105$),
which was originally classified as an XRF (Watson \etal\ 2004) based
on its X-ray brightness as inferred from the discovery of a halo from
dust scattering in our own Galaxy (Vaughan \etal\ 2004). However, an
analysis of its high energy spectrum recorded with {\it INTEGRAL}
(Sazonov \etal\ 2004) revealed an $E_p\sim190$ keV, apparently inconsistent with
that inferred from the dust halo. The paucity of distance measurements
thus prevents us from deriving definitive conclusions on the XRF
distance distribution.

 The spectroscopic signatures of type Ic supernovae seen in GRB~030329/SN~2003dh
 (Stanek \etal\ 2003; Hjorth \etal\ 2003), and the discovery of
 photometric bumps in many GRB optical afterglow light curves
 (e.g. Zeh, Klose \& Hartmann 2004), strongly support the collapsar
 model (Woosley 1993), whereby the majority of long duration GRBs are
 the result of core collapse supernovae. Should XRFs be associated
 with supernovae, as seen in GRBs, but found at typically lower redshifts, it
 should be possible to constrain their distances by the detection of a
 supernova brightening in their X-ray located host galaxies. This can be
 the case where an optical afterglow is seen (e.g. XRF 030723; Fynbo \etal\ 2004)
 or in some cases where there was no apparent optical afterglow 
  (e.g.,
 this was recently the case for GRB/XRF 031203 (Bersier \etal\ 2004;
 Thomsen \etal\ 2004;Cobb \etal\ 2004; Gal-Yam \etal\ 2004; Malesani \etal\ 2004)). 

Here we present results on our {\it HST} data of two XRF host
galaxies, identified earlier by a combination of {\it HST} and {\it
Chandra} data (Fruchter \etal\ 2002a,b; Bloom \etal\ 2003). In the
present study, we concentrate on the difference imaging between two
{\it HST} epochs separated by six months, and the constraints these
results place on underlying supernovae. Finally we combine these results with
the properties of the X-ray lightcurves of each XRF and their similarities with
GRBs to reach conclusions on the likely nature of XRFs.

\section{Observations}

XRF~011030 was detected with {\it BeppoSAX} on 2001 October 30,
06:28:02 UT (Gandolfi 2001) within an initial error circle of
$r=5$\arcmin. It had a long duration ($t_{90}\sim 1000s$; for a
definition of $t_{90}$ see Kouveliotou \etal\ 1993), an apparently low
$E_p$ ($<$ 40 keV; Heise \etal\ 2001b) and a fluence of $9 \times
10^{-7}$ ergs cm$^{-2}$ ($2-28$ keV). We imaged the region surrounding
XRF~011030 with the WIYN 3.5m telescope on 2001 November 1, 2, and 3,
reaching limiting magnitudes of R$\sim$ 23.5 on each occasion.
Image subtraction failed to locate any afterglow candidates to
these limits; a stacked image indicated that no optical afterglow was
present to R$>$24 (Rhoads \etal\ 2001). Likewise, other optical
observations taken by several groups also failed to identify any
afterglow candidates. Radio observations with the VLA, however, did
identify a fading source as the possible afterglow of XRF~011030
(Taylor \etal\ 2001).  The field was imaged by {\it Chandra} at two
epochs on 2001 November 9 (47.21 ks) and 29 (20.12 ks), with the
Advanced CCD Imaging Spectrometer S-array (ACIS-S). Comparison of the
two epochs revealed a variable source within the {\it BeppoSAX} error
box at  at RA =20:45:36.00, DEC =+77:06:01.09 (Harrison \etal\ 2001;
Bloom \etal\ 2003),
approximately 1\farcs 2 from the location of the radio
source; the \chandra\ source was then identified as the X-ray
counterpart of XRF 011030 (Fruchter \etal\ 2002a).

XRF~020427 was detected also with {\it BeppoSAX} on 2002 April 27,
3:48:40 UT and was classified as an XRF by its $10-28$ to $2-10$ keV
hardness ratio, which indicated an X-ray rich event (in't Zand \etal\
2002). Optical observations taken after the burst failed to 
locate the afterglow to a limit of I=21.7, 51 hours after the burst
(Jakobsson \etal\ 2004) .We observed the field of XRF 020427 on 2002 May 11 with the
1.54m Danish Telescope at La Silla, Chile in the R-band with a total
exposure time of $9 \times$ 300 seconds (Table 1) and place a limit
of I$>$22.0 for any optical afterglow at this time. Radio observations of the XRF field also
failed to find any counterpart (Wieringa \etal\ 2002). XRF~020427 was
imaged with the \chandra\ ACIS-S array on 2002 May 6 (13.92 ks) and 14
(14.76 ks). Comparison of the two observations revealed a bright,
rapidly declining source within the {\it BeppoSAX} error circle at RA =
22:09:28.22, Dec=-65:19:32.03 (Fox 2002a,b; Bloom \etal\ 2003). VLT observations of
the counterpart obtained in June 2002 show a blue colour complex of
galaxies underlying the position of the X-ray transient (Castro-Tirado \etal\
2002). Amati \etal\ (2004) presented a comprehensive analysis of all
available X-ray data on this burst. The prompt event had an $E_p < 5$
keV and a duration $t_{90} \sim 60$s, and its fluence was $5.8 \times
10^{-7}$ ergs cm$^{-2}$ ($2-28$ keV); the best fitting pseudo-redshift
was $0.1 < z < 0.2$ (Atteia 2003; see also section 4).

For each XRF we reprocessed the {\it Chandra} data, extracted the XRF
spectra and generated a response using {\it CIAO} version 3.0.2 and
{\it CALDB} version 2.26. We searched both images in a region
4.2\arcmin$\times$4.2\arcmin\ (512 $\times$ 512 pixels) centered at the
{\it BeppoSAX} source location; within this radius, there is only one
clearly decaying source, already identified as the XRF afterglow. We
used a source extraction region of 2\arcsec\ radius centered on each
XRF location to extract spectra in both epochs, and we fit the spectra
simultaneously with an absorbed power law model. For each burst we
used the C-statistic (appropriate for low count data; Cash \etal\ 1979); the
spectra were unbinned and the (negligible) background was not
subtracted. The resulting counts, spectral indices and fluxes are
shown in Table 1.

In the following we use our WIYN and Danish 1.5m images of XRF 011130
and 020427, respectively, to facilitate the alignment of the {\it
Chandra} observations to the {\it HST} field.
  
\section{{\it HST} Observations}

XRF 011030 was observed with {\it HST} on 2001 December 12 and again
on 2002 June 12. The first of STIS observations were obtained in both
the 50CCD and Long-Pass (LP) filters, while the second was obtained
only with 50CCD. The field of XRF~020427 was imaged with {\it HST}
also using STIS in the 50CCD mode 2002 June 10 and 14; LP images were
also obtained at the June 14 epoch. A final epoch was obtained on 2002
October 26, again using the 50CCD filter. Table 2 shows the log of the \hst\ observations. Both XRF
datasets were retrieved from the {\it HST}
archive\footnote{http://archive.stsci.edu} and were combined and
cosmic ray cleaned using the drizzle routine (Fruchter \& Hook
2002). Dithered images were drizzled onto a grid with pixels of size
0\farcs 025, half the native STIS pixel, using a value for {\tt
pixfrac} of 0.7. All observations were aligned with the IRAF routine
{\tt geomap} using point sources in common in each \hst\ image and
then redrizzled onto an aligned output grid. A log of {\it HST} and
all ground based optical observations is shown in Table 2.

\subsection{Astrometric Alignment}

To accurately position the {\it Chandra} sources on the {\it HST}
image the {\it Chandra} field has to be precisely aligned with that of
{\it HST}. The STIS FOV of only 50\arcsec\ typically contains an
insufficient number of sources to assure such an alignment and,
therefore, it is necessary to use ground based intermediate
images. Here we used data obtained at the WIYN (XRF~011030) and Danish
1.5m (XRF~020427) telescopes. Each of these fields was initially
aligned to the USNO-A2 catalog and positions were estimated for all objects for which an apparent optical counterpart was identified within 0\farcs 5 of a \chandra\ source. 
A first alignment of the X-ray to the optical sky was then performed using the \chandra\ header coordinates.
Finally, we performed relative astrometry between our ground based observations and the HST images, using eight and ten common point sources between each field (for XRF 011030 and
020427, respectively). As a result, we could place the X-ray
counterparts of XRF~011030 and 020427 onto our {\it HST} images with a
positional accuracy of $\sim$ 0\farcs2 and 0\farcs15,
respectively.

Our positions agree with those reported by Fruchter \etal\ (2002a,b)
and Bloom \etal\ (2003) and confirm their identification of the host
galaxies. The positions of the X-ray afterglows for both XRFs are
coincident with the stellar fields of their host galaxies with blue
global colors (Bloom \etal\ 2003). The host of XRF~020427 also appears
to have several neighbour galaxies of similar magnitude, color and
morphology indicating that it may be part of a galaxy group. The
blue colors of each XRF host galaxy are similar to those of other GRB
hosts, and imply relatively young stellar populations and little dust
(see e.g. Trentham \etal\ 2002; Le Floc'h \etal\ 2003; Christensen \etal\ 2004).

To search for optical variability of the X-ray sources, we performed
direct image subtractions between the \hst\ epochs for each XRF (two
for XRF 011030 and three for XRF 020427). To increase the depth of our
final subtraction for XRF~020427 we also drizzled the data of June 10
and 14 together onto a single first epoch image. We then searched the
resulting difference image using SExtractor (Bertin \& Arnouts 1998)
with a signal to noise threshold of 5$\sigma$ and a
Gaussian mask with FWHM = 3 pixels, to mimic the {\it STIS} point
spread function (PSF); no variable object was found in
either field, with the exception of saturated stellar objects, which
leave clearly detectable residuals. Our subtractions are shown in
Figure 1. To define the depth of our subtractions, we created
artificial stars within the first epoch image with magnitudes in the
range $27 < V< 30$, and a FWHM equal to the PSF in our {\it STIS} images. We
then performed the subtraction of the final from the initial epoch and
searched the resulting difference image for variable sources using the
same method described above. We recovered 100\% of the artificial
stars at $V = 28.7$ in the XRF~020427 image, and 95\% of those at $V=28.5$ in the
XRF 011030 field (at a 5$\sigma$ detection level). 

This sensitivity difference is mostly due to a larger number of saturated objects in
the field of XRF~011030 than for XRF~020427; when our artificial stars
land on (or near) these objects, we are unable to recover them. Moreover, 
the depth of XRF~011030 images is slightly lower. The ``true'' limit for XRF~011030 further diminishes
to $V\sim27.6$ when the large foreground extinction (E(B-V) = 0.4) towards this source is considered. For XRF~020427 the smaller extinction (E(B-V)=0.03) has little effect on the limiting magnitude. 
We discuss the limits these observations place on the optical afterglow
and underlying supernova emission in the next section.

\subsection{Lack of residual optical emission at 40 days after the burst trigger}

Assuming that there is a supernova associated with each XRF, the flux
seen at the location of any burst at a given time is the sum of the
afterglow, supernova and host galaxy. The host galaxy component can be
removed by the subtraction of a late time image after both afterglow
and SN have faded. For GRBs with bright optical afterglows their
magnitude at $\delta t = 10$ days after trigger, is in the range $22 <
R <25$ (see e.g. Fig 2. Fox et al. 2003); assuming a fast decaying
trend ($t^{-2}$) this would imply that the majority of these afterglows
would be somewhat brighter than $R=28$ after 40 days. In
the case of XRF~011030 and 020427 the putative sum of afterglow $+$ SN
at 40 days is fainter than $V \sim 28.5$. This is not overly
surprising however, since many GRB afterglows are often not seen at
optical wavelengths (so called ``dark'' bursts), and therefore a more careful
examination is necessary to determine if this lack of emission is
unusual.

Under the standard fireball model (see M\'esz\'aros 2002 for a review)
GRB afterglows are described by a synchrotron spectrum, which can be represented
to first order by a set of gradual power-laws with breaks at
frequencies representing the cooling of the fireball ($\nu_c$), the
peak frequency of the electrons $(\nu_p)$ and the self-absorption
frequency where the fireball becomes optically thick ($\nu_{sa}$). It
is therefore possible to extrapolate between wavebands under the
premises of this model. Such extrapolations are potentially of great
value since they predict the expected optical flux in a GRB and may
differentiate between truly dark events and the ones which are not
seen due to insufficient depth of the observations (Fynbo \etal\
2001). Groot \etal\ (1998) first attempted such a method for the dark
GRB~970828 and found that extrapolating the X-ray slope into the
optical regime predicted significant optical flux, while none was
seen. However, such predictions need to take account of the possible
presence of $\nu_c$ between the optical and X-ray frequencies and
therefore a range of possible slopes becomes necessary; thus the
distinction between dark and bright bursts is not sharp but is
populated with ``gray'' bursts, events which fall between the extremes
of the extrapolations (Rol \etal\ 2004). Applying the same method of
extrapolation to XRF~011030 and 020427, we find that the existing
optical limits for both bursts lie below the most optimistic estimates from
the X-ray afterglow of each burst, but above the limits derived using
a worst case scenario (the most rapid possible decay and the presence
of a cooling break between optical and X-ray bands). Thus these
non-detections can be explained without recourse to additional
extinction (and always assuming a fireball spherical expansion); this
is evident in Figure 2, where we show that the observed optical
limiting flux for XRF 020427 falls well within the extrapolation of
the limits obtained from the X-rays. For more details of
this method, the reader is referred to Rol \etal\ (2004).

We now proceed to examine the frequency of SN detection in GRB
afterglows. Since the discovery of the first such event
(GRB~980425/SN~1998bw; Galama \etal\ 1998) many low redshift ($z <1$)
afterglows have been found with indications of supernovae signatures
in their light curves (e.g. Zeh, Klose \& Hartmann 2004). These supernova
are thought to be the results of core collapse of massive stars (e.g.,
Wolf-Rayet stars) that have lost much of their hydrogen (and possibly
helium) envelopes prior to collapse and therefore form Type Ib/c supernovae
(Woosley 1993; MacFadyen \& Woosley 2000). The recent spectroscopic discovery of supernovae
associated GRBs in 2003 (GRB~030329; Hjorth \etal\ 2003; Stanek \etal\
2003, GRB~031203; Malesani \etal\ 2004) strongly supports this model.
An observational consequence of a GRB-SN association
is that, since supernovae reach their maximum light a few weeks after the core
collapse (which is thought to trigger the GRB), their emergence in the
lightcurve can slow or even reverse the fading of the optical
afterglow. More importantly, in some cases they may also be visible
even where an optical afterglow is not (e.g. SN~1998bw/GRB~980425;
SN2003lw/ GRB/XRF~031203; Galama \etal\ 1998; Bersier \etal\ 2004;
Cobb \etal\ 2004; Gal-Yam \etal\ 2004; Malesani \etal\ 2004; Thomsen
\etal\ 2004).

To determine the limits on underlying supernovae for each XRF, we therefore
created spectral templates based on the SN~1998bw spectra first
reported by Patat \etal\ (2001)\footnote{These templates were taken
from the suspect SN database
http://tor.nhn.ou.edu/\~~suspect/index.html}. Since our
observations at a fixed observer time correspond to different rest
frame times, for each redshift considered we chose the template
spectra taken closest to the rest frame time corresponding to the
observer time of the {\it HST} data.  In
cases where temporally coincident spectra were not available, we used
the closest available spectra normalized against the observed
photometry of SN~1998bw at that time (using the light curves of Galama
\etal\ 1998) and K-corrected them to the appropriate redshift
corresponding to the fixed observer time. We then determined the
measured counts from these spectra within the STIS/50CCD passband by
convolving them with the STIS response with {\it synphot}. At this
stage we also folded in the effect of Galactic extinction for each XRF
field based on the E(B-V) from Schlegel, Finkbeiner \& Davis
(1998). This allows us to obtain the expected number of counts from a
SN~1998bw like supernova for each XRF field at various redshifts.

The optical spectra available for SN~1998bw extend only to $\sim
3000$\AA. Therefore they are of limited use at higher redshifts, where
we observe the rest frame ultraviolet component. Moreover, very few studies exist to date
of the UV-spectra of type Ic supernovae, in particular temporally
resolved studies, which would be ideal for our comparisons (a generic
feature of most supernovae is a supression of the UV-flux due to the
blanketing effect of metals). We have
attempted therefore, to estimate limits on the UV flux via two
methods. The first is to assume zero flux below 3000\AA\ (rest frame);
while this is clearly underestimating the true flux, it provides a
firm lower limit on the expected brightness of the supernova within our
redshift limits. The second method is to use the available UV
spectroscopy of SN~2002ap (a typical Ic supernova), taken with {\it STIS} on 2002
February 1 using the G230L grism. This spectroscopy shows a substantial decrease in
flux from $\sim 3000-2000$ \AA\ as expected and also seen in SN~1994I
(Millard \etal\ 1999). We have rescaled the flux here for our purposes
so that it matched the flux seen with ground based spectra at $\sim
3000$\AA. Although these spectra should provide a reasonable measure
for the UV flux for type Ic supernovae, we also note that the spectrum of
SN~2002ap was taken at an early epoch ($\sim$ 5 days) and possible
evolution of the features may change the spectral shape at later
epochs. However, as time resolved UV spectroscopy is not available for
any high velocity Type Ic, we cannot determine if this early shape is maintained
into later epochs.

SN~1998bw was a very bright Type Ic event; fainter Type Ic supernovae are more
common locally and it is likely that their overall distribution may be
bimodal (Richardson \etal\ 2002). Although it is not clear that all Type Ic supernovae
produce a GRB, the ones firmly (i.e. spectroscopically) associated with GRBs are comparable
in brightness to SN~1998bw (e.g. Hjorth \etal\ 2003; Stanek \etal\
2003; Garnavich \etal\ 2003). The lack, however, of a supernova in GRB
010921 to a limit of 1.34 magnitudes fainter than SN 1998bw (Price
\etal\ 2003) and the best fit of both GRB 020410 and XRF~030723 with 
a supernova $\sim$ 2 magnitudes fainter than SN~1998bw (Levan \etal\ 2004a; Fynbo \etal\ 2004) 
imply that we may be seeing a broader 
luminosity function for the GRB associated supernovae. Low redshift $(z<1)$ GRBs show
a moderate ($\sim$ 1 magnitude) dispersion in the peak luminosity of their supernovae
(Thomsen \etal\ 2004; Zeh, Klose, \& Hartmann 2004).
The faintest of the local high-velocity type Ic supernovae is
SN~2002ap. This supernova exhibits similar spectral and temporal 
evolution to SN~1998bw (with a slightly faster rising lightcurve, e.g. Fig 4 of
Foley \etal. 2003) and was 2 magnitudes fainter than SN~1998bw although still
one magnitude brighter than the faintest type Ic ever seen (Richardson \etal\ 2002).
We have, therefore
chosen to use this supernova as a template to study what might be expected as a limiting magnitude
from fainter supernovae. Our predicted magnitudes are not strongly dependent
on the UV-flux below $z \sim 1$ as can be seen in Figures 3 \& 4\ (upper
panels), where we plot the evolution of the magnitude of SN~1998bw and
SN~2002ap at various redshifts in the two XRF fields,
respectively. However the limiting magnitude beyond this is a
sensitive function of the unknown UV-flux, and the true magnitude may
be somewhat different; a hard upper limit is set by our assumption of
zero UV flux.

The evolution of the expected magnitudes for supernovae with differing 
properties are shown in Figures 3 \& 4 (for XRF 011030 and 020427 respectively).
For XRF~011030 a faint type Ic supernova association, such as SN~2002ap,
would fall below the detection limit of our observations at $z \sim 1$,
while a SN~1998bw type would have been seen out to $z \sim 1.5$. For
XRF~020427 both supernovae would be visible up to $z \sim 1.2$, and
again only a SN~1998bw type could be seen beyond $z\sim 1.5$. However,
for $z>1$ the rest frame UV light from the supernova is very sensitive to the
host galaxy extinction, such that a moderate $A_V \sim 1$ could
extinguish the supernova contribution (see also Figures 3,4, lower
panels). Further, the true colors of type Ic supernovae are not well
constrained in the rest frame UV. For most supernovae the colors evolve from
blue to red, with bluer bands experiencing peak light before the
red. Some supernovae associated with GRBs, however, have exhibited
blue (rest-frame UV) peak colors (e.g., SN2001ke/GRB~011121; Garnavich
\etal\ 2003; Greiner \etal\ 2003) implying that they have a
significant UV-component, while many Type Ic supernovae (including SN~1994I
and SN~2002ap) have a large UV decrement due to metal line
blanketing. Our approach, therefore places only a lower limit on the
expected emission from supernovae at these redshifts.

The limiting magnitudes placed on underlying supernova depend not
only on the observed spectra but also on the shape of the light curve of
the associated supernova. Local type Ib/c supernova show a range of
light curve properties including some which evolve faster than SN~1998bw
(e.g. SN~1994I) and some which evolve slower (e.g. SN~1997ef). 
A fast evolving supernova would have decayed further at the time of our 
observations, while a slower one would be closer to peak 
(or even at peak if viewed at higher redshifts). To explore the possible effect of 
light curve shape on our limits we performed the same K-corrections
as described above but on supernovae which evolved 30 \% slower, and
30 \% faster than SN~1998bw. In each case we maintained the UV-flux 
estimates for SN~2002ap. A faster evolving supernova would drop 
below our visibility limit at $z \sim 0.8$ for XRF~011030 and at $z \sim 1.1$ for
XRF~020427, assuming the same peak magnitude with SN~2002ap at $M_V = -17.2$. A slower evolving supernova
(again with a peak brightness similar to the one of SN~2002ap) would be seen out to $z=1.3$ and $z=1.5$ for XRF~011030 and 020427, respectively. It should also be noticed that our estimates
assume that the supernova light at the time of our second epoch observations is much fainter than
at the first. This is so far the case for {\it all} type Ib/c SNe known (for example SN~1998bw would have fallen by more than a factor 10 in the period between the observations considered here); it is clear, however, that our estimates depend on the SNe light curve shape and would provide different limits in the (unlikely) event of e.g., SNe light curves exhibiting a plateau. 

\section{Discussion}

The X-ray light curve decay indices (assuming a power law decay
starting at $t=0$ for each burst) are $\alpha = -2.0 \pm 0.3$ and
$\alpha = -1.4 \pm 0.4$ for XRF 011030, 020427, respectively. These
slopes are both comparable to those observed in typical GRB X-Ray
afterglows, which at late times lie in the range $-1 < \alpha < -2$ with
a best fit of $\alpha ={-1.69}$ (Kouveliotou
\etal\ 2004). Unfortunately our sample of known redshift XRF
afterglows with multi-epoch X-ray imaging is very limited with a
possible exception of the still debated GRB/XRF~031203, which exhibits
a large X-ray flux as well as a high peak energy (Vaughan \etal\ 2004;
Watson \etal\ 2004; Sazonov \etal\ 2004). The fluxes observed for 
XRF 011030 and 020427 are also similar to the typical GRB fluxes at $z\sim1$ at similar
epochs. For either XRF afterglow, however, to have a luminosity
similar to GRB~980425 or GRB/XRF~031203 they would have to be
exceptionally close ($z \sim 0.1$); conversely, when placed at
redshifts higher than $z \sim 1.5$, they would become among the most
intrinsically luminous afterglows yet observed (Kouveliotou \etal\ 2004).

It is interesting to investigate how off-axis models apply to
XRFs. Off-axis models predict that the early X-ray light curves will
differ significantly from those of on-axis events. Specifically the light curves should
start substantially fainter and then rise as the core of the on-axis
material slows down and comes into the observer's line of sight. An
observer at $\theta_{\rm obs} > \theta_0$, where $\theta_0$ is the
initial jet opening angle, sees a rising light curve at early times,
peaking when the jet Lorentz factor is $\sim 1/\theta_{\rm obs}$, and
approaching that seen by an on-axis observer, at later
times. Therefore under this model we would expect to find that at the
time of our X-ray observations the XRF afterglow luminosities, if
located at $z>1$, should be smaller than those of classical, on-axis
GRBs. For XRF~020427 early {\it BeppoSAX} observations imply that the
afterglow decayed approximately monotonically from an early stage, and
thus is incompatible with a sharp edge seen off-axis (see
e.g. Ramirez-Ruiz \& Madau 2004); the lack of early X-ray data for
XRF~011030 prevents us from drawing quantitative conclusions. Another
possibility for XRF~020427 is that the jet does not have sharp edges
but wings of lower energy and Lorentz factor that extend to large
$\theta$. Such a picture of the jet is consistent with the
relativistic studies of the collapsar model by Zhang, Woosley, \&
MacFadyen (2003) and Ramirez-Ruiz, Celotti \& Rees (2002). The
early emission detected by {\it BeppoSAX} in XRF~020427 would be then
produced by material (on the wings) moving in our direction.

For a dirty fireball or an inefficient shock model the degree of
collimation can be similar to that of a GRB, and an XRF can still be
viewed directly down the jet-axis. The evolution of the XRF light
curve could be similar to that of a classical GRB. Unfortunately the sparse
multiwavelength afterglow coverage makes any quantitative analysis in
this case unreliable. For XRF~011030, however, the observed modulation during the first
$\sim14$ days in the 4.86 GHz and 8.46 GHz
data\footnote{http://www.aoc.nrao.edu/$\sim$dfrail/011030.dat}, and the
swings in the radio spectral index, bear the signature of diffractive
scintillation. If this is indeed the case then the angular size of
the source must be less than the diffractive angle $\theta_d$. For
typical parameters and an observing frequency of 8.46 GHz,
$\theta_d=3$ microarcseconds.  For a relativistic expanding fireball,
the linear size of the source is $R=f\gamma^2ct$ and the apparent size
is $R/\gamma$ (here $\gamma$ is the Lorentz factor for the expanding
shell, and the factor $f$ depends on the dynamical details of the
expanding fireball model (see e.g. Katz \& Piran 1998)). Thus, the
apparent size of the source is $f\gamma ct \le 2\times 10^{17} (f/4)
(\gamma/2) (t/2{\rm \;weeks})$ cm. Taking the above-mentioned size of 3
microarcseconds for the fireball at about $t=$2 weeks post-burst, we
can obtain a constraint on the distance to the source. Using a value
of 72 km s$^{-1}$ Mpc$^{-1}$ for the Hubble constant, we find $z
\gtrsim 0.25$.

GRB host galaxies show a broad range of absolute magnitudes and 
physical extents (e.g. Hogg \& Fruchter 1999; Bloom, Kulkarni \& Djorgovski 2002).
For both XRF~011030 and 020427 the
angular size and absolute magnitude lie within the range seen for
GRB host galaxies across a broad range of redshifts ($0.1 < z < 2.5$).
Thus we cannot use morphological or luminosity information 
to constrain the distances to these galaxies.

We post below several explanations by which the lack of a supernova
component can be accounted for. The first option is that XRF~011030 and 020427 are accompanied by bright supernovae but lie at $z>1.2$-$1.5$. At this redshift XRF~020427 would clearly violate the observed $E_p$-$E_{iso}$ relationship reported by Amati \etal\ (2002), Lloyd-Ronning \& Ramirez-Ruiz (2002) and subsequently Lamb \etal\ (2004) using data from {\it BeppoSAX}, {\it BATSE} and
{\it HETE-II}, respectively. This relationship finds that in GRBs, $E_{p}
\propto E_{iso}^{1/2}$. Under this scheme the isotropic energy of
XRFs must be lower than that seen for GRBs, which would in turn result
in a shorter distance distribution.  To date it has been difficult to extend this
relation into the XRF regime (especially at very low $E_p <10$ keV) since
only one XRF has a firmly established redshift (XRF~020903 at $z=0.25$;
Soderberg \etal\ 2003). However the existence of GRB/XRF~030429 at
$z=2.66$ (Jakobsson \etal\ 2004; Sakamoto \etal\ 2004) implies that some 
outliers do exist, although the $E_p$ = 35 keV for
this burst is not sufficiently low to be greatly constraining. Likewse the $E_p$ limit
on XRF 011030 of $E_p < 40$ keV does not provide a strong indicator of
redshift. However for XRF~020427 the $E_p < 5$ keV does impose stronger limits. 
Amati \etal\ (2004) derive a best fit redshift range of $0.1 < z < 0.2$ for XRF~020427 and
they notice that at the maximum allowed redshift ($z<2.3$ based on
the optical spectra of van Dokkum \& Bloom 2003) XRF~020427 would not fit
the $E_p$-$E_{iso}$ relation since this would require $E_p > 100$ keV.
We also find that if at $z>0.5$, XRF~020427 lies outside the 90 \% bounds of the Amati relation, assuming the latter is described by $E_{p} = 89 (E_{iso} / 10^{52})^{0.5}$ with a lognormal scatter of 0.3 (Butler \etal\ 2004). 
Notably GRB~980425, also violates this relationship, with a very low
$E_{iso}$ and high $E_p$. In contrast, should XRF~020427 be at higher-$z$, it would be the most extreme of the outliers with low $E_p$ and high $E_{iso}$.

Another straightforward explanation would be that these XRFs were 
not accompanied by a luminous supernova. While
this is possible, it is puzzling given the clear observational
evidence for supernovae now seen in many low redshift,  long duration GRBs, 
the possible association of an XRF with a supernova (XRF~030723; Fynbo
\etal\ 2004) and the spectroscopic supernova signatures seen in
GRB/XRF~031203.

Further possibilities which can explain {\it both} a low redshift and the 
absence of a supernova are, (i) that the afterglows were heavily
dust extinguished, (ii) that the associated supernovae were very faint,
or (iii) that the supernova and XRF were not temporally coincident in 
each of these cases. We now consider each of these options in turn.

In the high extinction scenario both afterglow and supernova would not be
seen due to the strong absorption of optical light. The colors of their host galaxies, however, imply that
they contain little dust. For XRF~020427, this is confirmed by deep
K-band imaging that rules out very red colors; the limit is formally $R-K <
2.5$ (Bloom \etal\ 2003). The caveat here is that these observations
do not probe the line of sight directly to the source of the burst,
which could potentially be heavily extinguished. To investigate this
possibility we have plotted in the bottom panel in Figures 3 \& 4 the
extinction necessary in the host galaxy of each XRF to hide a SN as
strong as SN~1998bw and SN~2002ap. Here we have assumed an SMC like
extinction law (Pei 1993), found to be the best fit to many GRB afterglows
(e.g. Jakobsson \etal\ 2004; Holland \etal\ 2003). Similarly, the
majority of GRBs do not show excessive extinction in their light
curves and only GRB~030115 ($A_V \sim 1.5$) Levan \etal\ , in prep; Lamb \etal\ 2003) has a
significant dust column. Thus if these two XRFs do demonstrate heavy
reddening, then they are very different from the optically bright
GRBs.

One way of determining the possible extinction along the
line of sight is to measure the hydrogen column density ($N_H$) from
the X-ray data. Each XRF has a relatively low number of counts (see
Table 2) and therefore the limits which we can place are unfortunately
poorly constrained. However in each case we find a value marginally in
excess (but consistent with) the Galactic value.  The 95 \% upper
limits on the value of $N_H$ (Galactic + host galaxy) are $\sim 3.5
\times 10^{21}$ cm$^{-2}$  and $\sim 4 \times 10^{21}$ cm$^{-2}$ for
XRF~011030 and XRF~020427, respectively.
Accounting for the Galactic column and using Bohlin \etal\
(1978) to convert from $N_H$ to $A_V$, this implies an upper limit on
the host galaxy extinction of $A_V < 1.7$ and $A_V <2.5$ for
XRF~011030 and XRF~020427, respectively.  These poorly constrained limits
are also consistent with zero host galaxy extinction; we note, however,
that the conversion between $N_H$ and $A_V$ is sensitive to the assumed 
dust to gas ratio and therefore somewhat uncertain.

Since the extinction required to remove a supernova signature at low redshift ($z<0.5$)
is very large and apparently incompatible with the measured dust columns it is logical 
to consider if a fainter supernova could explain these observations. 
Richardson \etal\ (2002) conducted a survey of the absolute magnitude distributions of all types of supernovae; they found that the faintest SN Ic peaked
at approximately $M_V = -16$. Assuming a similar light curve evolution
to SN~1998bw, such a SN would have been visible at $z \sim 1$ with no
host galaxy extinction. As the supernova peak light scales linearly with its Ni yield, we would expect very little Ni production from a very faint supernova. The faintest local core collapse supernovae are of type II-P (again see Richardson \etal\ 2002), which approach $M_V \sim -14$; such supernovae would only be seen at low redshifts ($z < 0.5$). It is, however, by no means clear that such supernovae are capable of producing GRBs.

A final option is that the supernova and XRF are not temporally
coincident. In the collapsar model (Woosley 1993) the GRB occurs within
seconds of the supernova explosion, while other models have suggested
that a longer delay could occur between the two events. For example in
the supranova model (Vietri and Stella 1998) the GRB can be 
produced months to years after the SN. Long delays are however not favored by
observations of low redshift GRBs which support simultaneous
explosions (Galama \etal\ 1998; Hjorth \etal\ 2003; Stanek \etal\
2003). At moderate redshifts, the supranova model may be a viable
explanation for the lack of supernovae associated with XRF 011030 and 020427,
as a supernova occuring a few weeks before the XRFs would not have
been detected; at very low redshifts ($z < 0.2$) the delay would have
to be over 6 months.

\section{Conclusions}

We have presented deep, multi-epoch \hst\ observations of the fields
of two XRFs. These observations, reaching depths of $V \sim 28.5$
would be sensitive enough to detect underlying supernovae such as
SN~1998bw or SN~2002ap out to $z >1$ even with moderate host galaxy
extinction. At very low redshifts the extinction required to remove
the supernova flux is very large, and incompatible with the measured
column density in the case of XRF~020427. At higher redshifts XRF 020427
does not fit the $E_p$ - $E_{iso}$ relationship. This situation can
be reconciled if the XRFs are situated locally behind large dust
absorbing columns, if the supernovae are very faint, or if the SN and XRF are not temporally
coincident. A final option is that GRBs and some XRFs do not represent the
same physical process, and that they may be due to similar but
physically distinct phenomena, where a supernova is not required for
the latter.

The recent XRF~040701 may allow greater insights into the XRF-SN association, or the lack thereof. Optical
observations failed to locate either a supernova or an optical
afterglow (Berger \etal\ 2004; Pian \etal\ 2004), while {\it Chandra}
observations did find a candidate X-ray afterglow (Fox \etal\ 2004)
and a coincident galaxy system at low redshift $z=0.22$. Deep {\it
HST} observations of this object may significantly aid our
understanding of the XRF phenomenon.

\section*{Acknowledgments}
Support for Proposal  GO 9074 was provided by NASA through a grant  from the Space Telescope Science Institute, which is operated by the Association of  Universities for Research in Astronomy, Incorporated, under NASA contract  NAS5-26555. This work is also based partly on observations made with the Chandra X-ray Observatory, and at the Wisconcin, Indiana, Yale, and National Optical 
Astronomy Observatories (WIYN), and the Danish 1.5m. telescope at La Silla, Chile under
ESO program 69.D-0701.
This work was conducted in part via collaboration within the the
Research and Training Network ``Gamma-Ray Bursts: An Enigma and a
Tool'', funded by the European Union under contract number
HPRN-CT-2002-00294. A.J.L. acknowledges receipt of a PPARC studentship
and support from the Space Telescope Science Institute Summer Student
Program.  S.K.P. acknowledges support from HST Grant Number GO 9074 and SAO
grant number GO1-2055X.  This work was also supported by the Danish Natural 
Science Research Council (SNF). We thank the referee for a prompt and constructive report.

\begin{figure}
\psfig{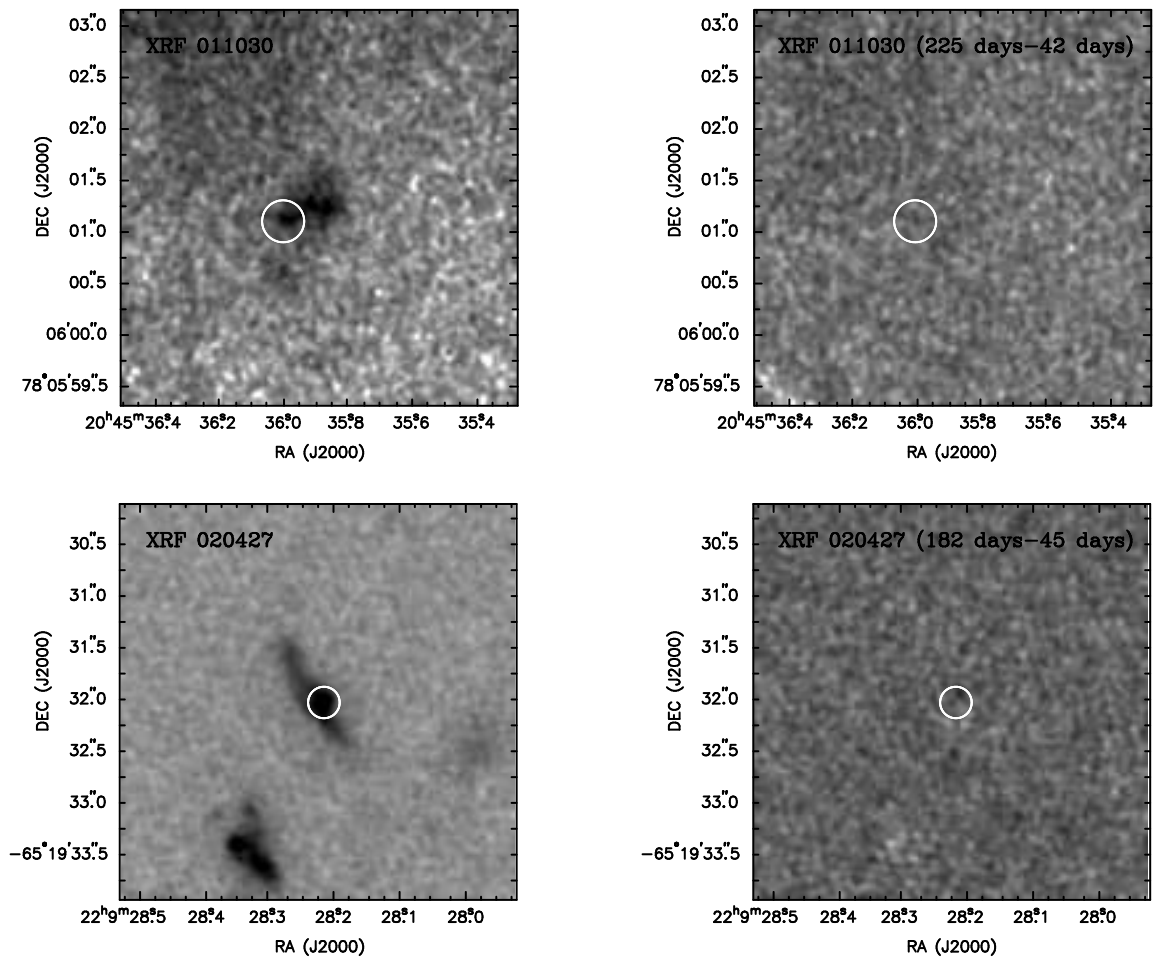}
\caption{The host galaxies of XRFs~011030 and 020427 (left column) and
the results of difference imaging on each of their fields (right
column). For each XRF the images were obtained with STIS.
The two subtracted images are taken approximately six months
apart and show no evidence for any excess emission at either epoch,
placing limits of $V \sim 28.5$ for the combination of afterglow and
supernovae approximately 40 days after each burst. In each image North is
up, and East to the left.}
\label{ab}
\end{figure}

\begin{figure}[h]
\begin{center}
\hbox{
\psfig{figure=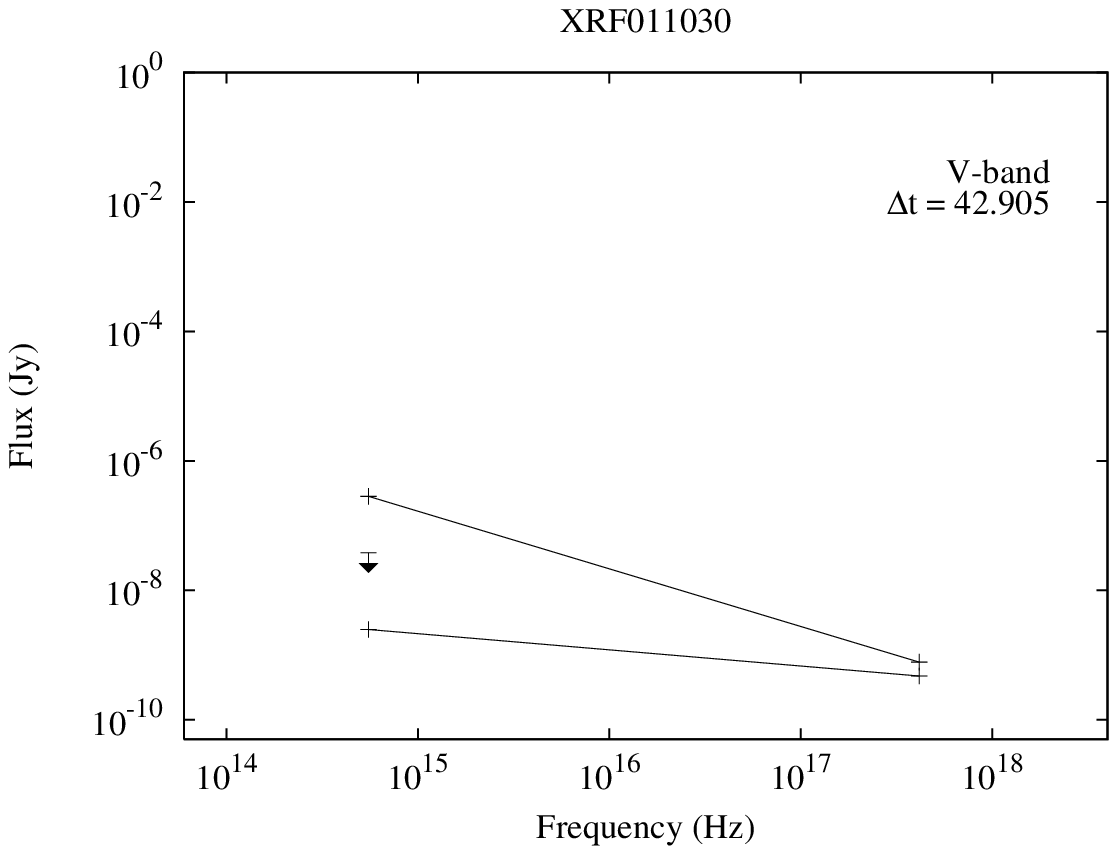,height=6.0cm,width=8.0cm}
\hspace{0.5cm}
\psfig{figure=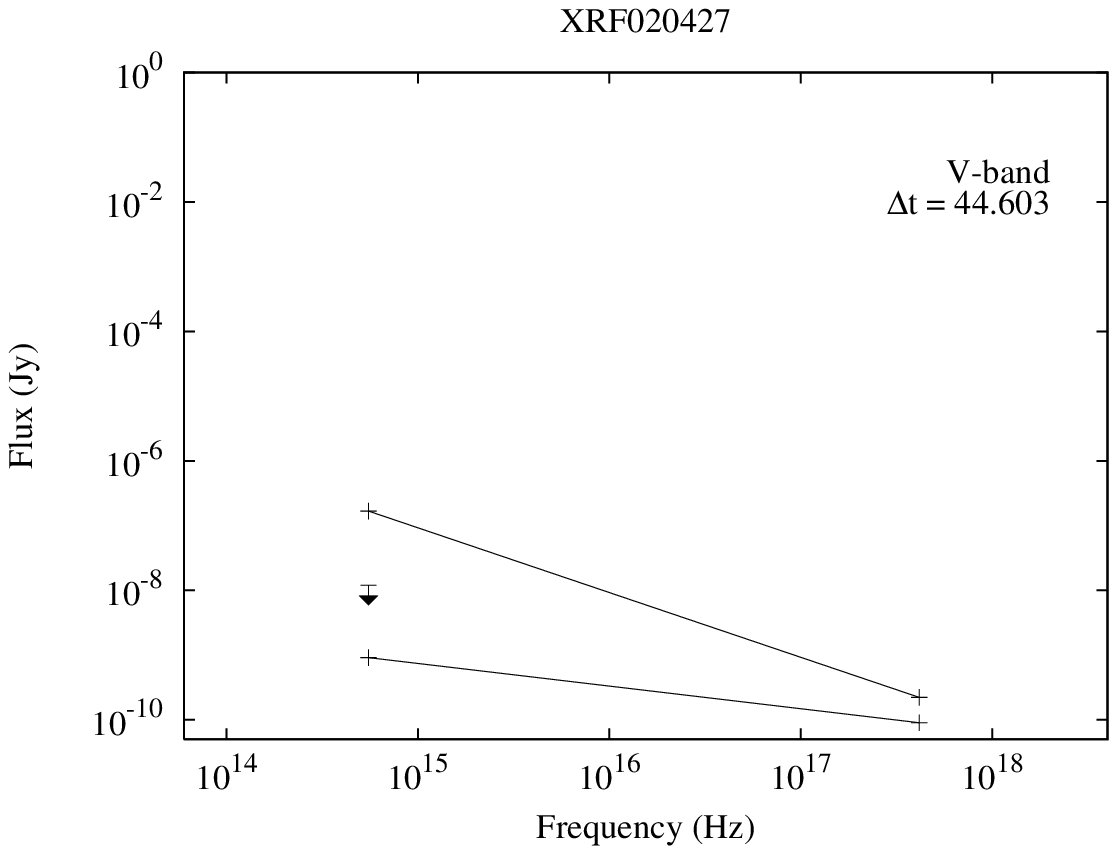,height=6.0cm,width=8.0cm}
}
\vspace{-1.0cm}
\end{center}
\caption{The extrapolation of the X-ray afterglows of XRF 011030 (left)
and XRF 020427 (right) into the optical waveband (STIS). The optical limits shown
are derived from extrapolating the X-ray observations both temporally and
spectrally to the optical observations. This was done using the
powerlaw indices from the X-ray observations, and allowing for a possible
cooling or jet break between the epoch of the X-ray and optical
observations, such that the final optical limits derived are the extremes
of all possible allowed scenario's (see Rol \etal 2004). }
\label{fig2}
\end{figure}

\begin{figure}
\begin{center}
\psfig{figure=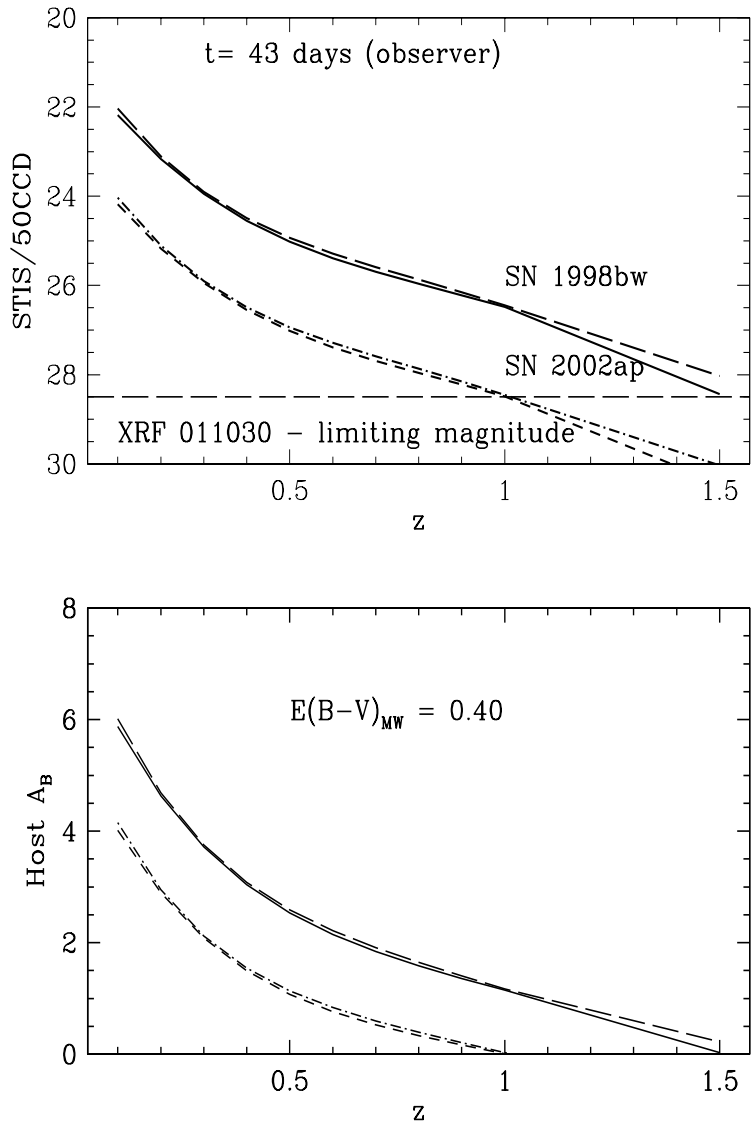,height=15cm,width=10cm}
\end{center}
\caption{{{\it upper panel}: Evolution of the magnitude of SN~1998bw
and SN~2002ap at various redshifts in the XRF~011030 field and at a
fixed observer time. Two lines are shown for each supernova, the lower line
(solid for SN~1998bw, dashed for SN~2002ap) show the extrapolations
assuming zero flux for $\lambda < 3000$\AA, while the upper line shows
the magnitudes assuming that the UV-flux had a spectral shape as
determined from SN~2002ap. The horizontal line corresponds to the
limiting XRF magnitude calculated in this paper. {\it bottom panel}:
The required rest frame B-band extinction within each XRF host galaxy
such that either a 1998bw or a 2002ap like SN could be hidden. The two
lines again represent different methods of treating the UV component
of the supernova flux.}}

\label{ab}
\end{figure}
\begin{figure}
\begin{center}
\psfig{figure=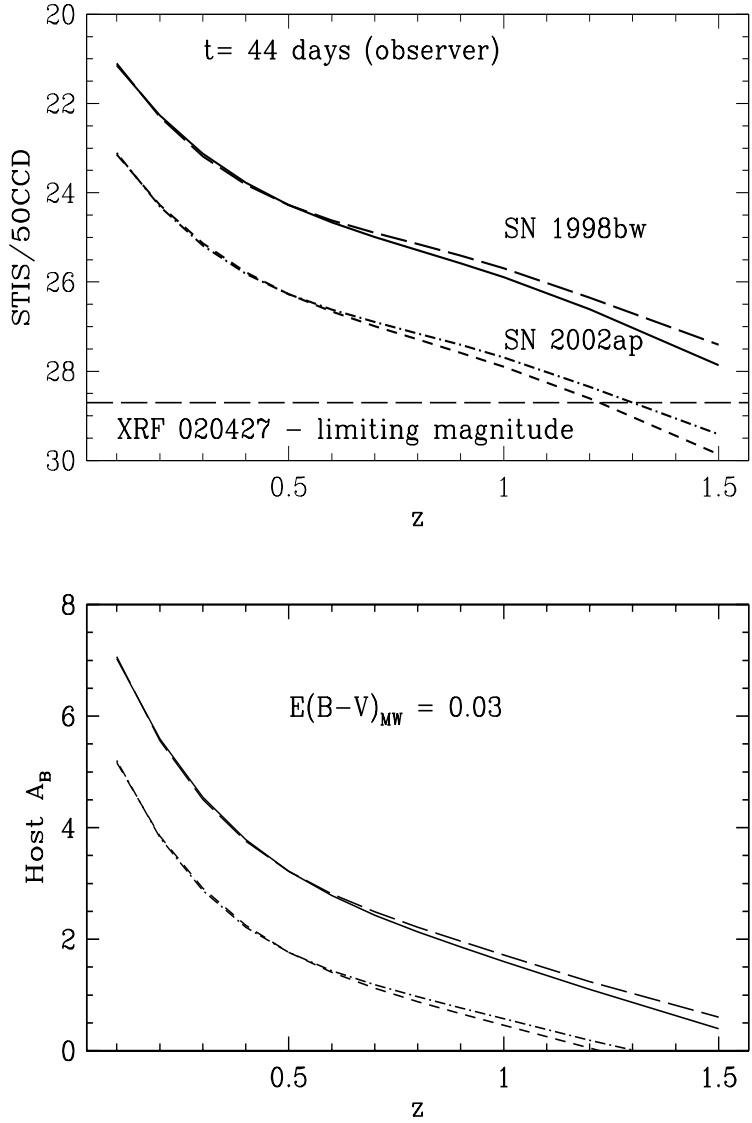,height=15cm,width=10cm}
\end{center}
\caption{{As for Figure 3, but for the XRF~020427 field. 
Note the limits shown in each figure (dashed line in the upper panel) are
the raw observed limits not corrected for Galactic extinction, since the extinction
is considered when the spectra are folded through the instrument response.
(i.e. the expected magnitude plotted in the upper panel is that expected for
the Galactic extinction relevant for each XRF).}}
\label{ab}
\end{figure}

\begin{deluxetable}{llcrccc }
\footnotesize
\tablecolumns{4}
\tablewidth{0pt}
\tablecaption{{\it Chandra} Observations of XRF~011030, XRF~020427}
\tablehead{\colhead{Date} & \colhead{$\Delta t^{b}$} &
\colhead{$t_{exp}$ (ks)} & \colhead{Counts$^{c}$} &  \colhead{$\Gamma$}
& \colhead{$N_{H}$ ($10^{21}$ cm$^{-2}$)} & \colhead{$F_x^{c}$}}
\startdata
XRF 011030\\
\hline
2001 Nov 09.728 & 10.46  & 46.6  & 374 & 1.31(8) & 1.02   & 1.06
$\times 10^{-13}$\\
2001 Nov 29.464 & 30.19  & 19.9  & 16  & 1.3(4)  & 1.02   & 1.07
$\times 10^{-14}$\\
\hline
XRF 020427 \\
\hline
2002 May 06.225 & 9.08  & 13.7 & 56 & 1.3(2) & 0.29   & 5.1 $\times
10^{-14}$\\
2002 May 14.154 & 17.00 & 14.6 & 24 & 2.1(3) & 0.29  & 1.2 $\times
10^{-14}$\\
\hline
\enddata
\tablecomments{$^{a}$ values in parenthesis correspond to 68\%
uncertainty, $^{b}$ days since burst,
$^{c}$ measured counts and unabsorbed flux (erg s$^{-1}$ cm$^{-2}$)
between $0.3-10$ keV, $^{d}$ Galactic
value (Dickey \& Lockman ,1990) ,
indicates that the spectral index is determined by a
joint fit to both data sets.}
\label{tab:photdata}
\end{deluxetable}

\begin{deluxetable}{llcrcc}
\footnotesize
\tablecolumns{6}
\tablewidth{0pt}
\tablecaption{Optical Observations of XRF~011030, XRF~020427 and 
their host galaxies}
\tablehead{\colhead{Date} & \colhead{$\Delta t = t-t_0$ (days)} & \colhead{Inst./Filter} &
\colhead{Exp. Time (s)} & 
\colhead{mag.}}
\startdata
XRF 011030 \\
\hline
2001 Nov 1.221&0.952	& WIYN/R        &3000  &$>22.5$ \\
2001 Nov 3.132&2.862	& WIYN/R	      &3000  &$>22.2$\\
2001 Dec 12.175&42.905	& STIS/50CCD  & 8640 &25.31 $\pm$ 0.10\\
2002 June 12.602 &225.380 	& STIS/50CCD  & 7505 & 25.30 $\pm$ 0.10\\
\hline
XRF 020427 \\ 
\hline
2002 May 11.607&14.447& Danish 1.54m/I & 2700 & $>22.0$\\ 
2002 June 10.762&44.603& STIS/50CCD  &8640 &24.44 $\pm 0.05$\\
2002 June 14.734&48.575& STIS/50CCD & 4781& 24.45 $\pm 0.05$\\
2002 Oct 26.011&181.852 & STIS/50CCD  &8392 &24.44 $\pm 0.05$ \\
\enddata
\tablecomments{Photometry measure is of the host galaxy. All photometry is corrected for galactic extinction. For R and I band observations the corrections are based on the values of
Schlegel, Finkbeiner \& Davis (1998). For the broad band 50CCD images we estimate the
extinction based on the change in count rate observed for a typical GRB host galaxy spectral
slope of $\nu^{-0.8}$ when folded through the 50CCD passband. The corresponds to $A_{50CCD} 
= 1.17$ and 0.06 for XRF~011030 and 020427 respectively.}
\label{tab:photdata}
\end{deluxetable}

\end{document}